\documentclass[12pt]{article}

\usepackage[T1]{fontenc}
\usepackage{graphicx}
\usepackage{a4wide}
\usepackage{textcomp}  
\usepackage{amsmath,amssymb}
\usepackage{units}
\usepackage{cite}
\usepackage[symbol]{footmisc}

\usepackage{titlesec}
\titleformat{\section}{\normalfont\bfseries}{\thesection.}{1em}{}
\titleformat{\subsection}{\normalfont\bfseries\itshape}{\thesubsection}{1em}{}
\titleformat{\subsubsection}[runin]{\normalfont\bfseries}{\thesubsubsection}{1em}{}

\renewcommand{\vec}[1]{\ensuremath{\mathbf{#1}}}

\renewcommand{\Im}{\ensuremath{\mathrm{Im}}}

\newcommand{\Lsys}{\ensuremath{\mathcal{L}}}
\newcommand{\Rsys}{\ensuremath{\mathcal{R}}}
\newcommand{\Bsys}{\ensuremath{\mathcal{B}}}
\begin{document}

%
%
\thispagestyle{empty}
\begin{center}
  \textbf{Spin-dependent tunneling in the nearly-free-electron model}

  \vspace*{2ex}

  P. BOSE$^{\star}$\dag, J. HENK\ddag \, and I. MERTIG\dag

  \vspace*{2ex}

  \dag  Martin-Luther-Universit\"at Halle-Wittenberg, Germany\\
  \ddag Max-Planck-Institut f\"ur Mikrostrukturphysik, Germany

  \vfill

  \begin{minipage}[t]{0.9\textwidth}
    \textit{Short title:} Computing tunnel conductances by NFE
  \end{minipage}

  \vspace*{5ex}

  \begin{minipage}[t]{0.9\textwidth}
    \textit{Words:} 5860

  \end{minipage}

  \vspace*{5ex}

  \begin{minipage}[t]{0.9\textwidth}
    $^{\star}$\textit{Corresponding author:}\\     
    P. Bose\\
    FB Theoretische Physik, FB Physik\\
    Martin-Luther-Universität Halle-Wittenberg
    D-06099 Halle (Saale), Germany\\
    Electronic address: bose@physik.uni-halle.de\\
    Telephone: +49 (0)345 5525-460\\
    Fax: +49 (0)345 5525-446

  \vspace*{2ex}

    Dr.\ J. Henk\\
    Max-Planck-Institut f\"ur Mikrostrukturphysik\\
    Weinberg~2, D-06120 Halle (Saale), Germany\\
    Electronic address: henk@mpi-halle.de\\
    Telephone: +49 (0)345 5582-970\\
    Fax: +49 (0)345 5582-765

  \vspace*{2ex}

    Prof.\ Dr.\ I. Mertig\\
    FB Theoretische Physik, FB Physik\\
    Martin-Luther-Universität Halle-Wittenberg
    D-06099 Halle (Saale), Germany\\
    Electronic address: mertig@physik.uni-halle.de\\
    Telephone: +49 (0)345 5525-430\\
    Fax: +49 (0)345 5525-446 
  \end{minipage}

\end{center}

%
%
\clearpage
\begin{center}
  \begin{minipage}[t]{0.9\textwidth}
    Spin-dependent ballistic transport through a tunnel barrier is
  treated within the one-dimensional nearly-free-electron model. The
  comparison with free electrons reveals significant effects of band
  gaps, in particular in the bias dependence. The results are
  qualitatively explained by the number of incident and transmitted
  states in the leads. With an extension to ferromagnetic leads the
  bias dependence of tunnel magneto-resistance is discussed.
  \end{minipage}
\end{center}

  \begin{minipage}[t]{0.9\textwidth}
    \textit{Keywords:} Nearly free electrons; Tunneling; Ballistic transport; 
                       Bias dependence; Tunnel magneto-resistance; 
  \end{minipage}

  \vspace*{2ex}

%
%
\clearpage

\section{Introduction}
\label{sec:introduction}
Spin-dependent transport of electrons in nanostructures is one of most
rapidly evolving areas of contemporary physics
\cite{Maekawa02,Zutic04}. Important contributions to the field come
from applied physics, focusing on design and characterization of
devices and applications, and from experimental physics, aiming at the
understanding of fundamental effects.  Contributions from theoretical
physics comprise on one hand transport calculations that are based on
state-of-the-art electronic-structure calculations. By this means,
properties of specific systems are investigated. On the other hand,
simple---and sometimes \emph{too} simple---model calculations are
performed in order to explain experimental results, thereby neglecting
occasionally important aspects of the system under consideration.
Apparently, there is a considerably large gap to be filled between the
advanced electronic-structure and the model calculations.

The giant magneto-resistance (GMR) had and still has a huge impact on
industrial applications \cite{Daughton99}. The tunnel
magneto-resistance (TMR), however, is expected to have even more
potential for future applications and devices
\cite{Grundler02,Klemmer02}. From a theoretical point of view, TMR has
the advantage that it can be understood in terms of simple
quantum-mechanical effects (e.\,g., the tunnel effect) and can be
modeled rather easily. Therefore, it lends itself support for an
investigation which connects model calculations to the frontiers of
nanoscience.

The purpose of the present paper is to provide a theory of
spin-dependent transport through tunnel junctions which bridges the
aforementioned gap. Its framework is the Landauer-B\"uttiker theory
\cite{Imry99}, in which elastic and ballistic transport is viewed as
transmission of scattering channels through the device.  All
ingredients needed for the calculations of the tunnel conductance can
be computed step-by-step, therefore allowing for
simplifications or extensions. The theory outlined here was
implemented one-to-one in a set of \textsc{Mathematica}\texttrademark\
notebooks \, \cite{Wolfram96}, in order to obtain numerical results.
Hence, the properties of the tunnel junction can be easily
manipulated, results for different set-ups computed and visualized
rapidly \cite{Mathematica}.

Model calculations typically assume free electrons in the metallic
leads that are connected to the tunnel barrier, whereas
electronic-structure calculations deal with a much more complicated
bandstructure. An important feature of the latter---which is missing
in the former---are band gaps, i.\,e., energetic regions in which
electronic states are not present.  Obviously, these have a
significant impact on the electronic transport.  In order to take band
gaps into account, we go beyond the free-electron approximation by
applying the nearly-free-electron (NFE) model. Assuming
one-dimensional leads, the theory is kept simple enough to be
tractable semi-analytically but it contains already the relevant
aspects of the electronic structure.  Both ferromagnetic and
nonmagnetic leads are discussed, because the results for the latter
can easily be transfered to the magnetic case. We note in passing that
tunnel junctions with ferromagnetic leads typically comprise
transition metals (Fe, Co, Ni), the electronic structure of which
cannot be described well within the free-electron model.

The paper is organized as follows. First, basic theories for
ballistic transport which are relevant to the current work are
sketched (section~\ref{sec:basic-theor-ball}). Having introduced the model
for the tunnel junction (section~\ref{sec:tunnel-junction}), the
electronic-structure calculations (section~\ref{sec:electronic-structure})
and the treatment of tunneling (section~\ref{sec:tunneling}) are presented.
Representative results of the tunneling calculations are discussed
in subsection~\ref{sec:ferromagnetic-leads-1}, focusing first on tunnel junctions with
nonmagnetic leads (subsection~\ref{sec:non-magnetic-leads-1}). Spin-dependent
transport is investigated in subsection~\ref{sec:ferromagnetic-leads-1}.

\section{Basic theories for ballistic transport}
\label{sec:basic-theor-ball}
The theory presented in the forthcoming sections applies to elastic
and ballistic transport of electrons through a tunnel junction
(figure~\ref{fig:tunbarr}). Hence, the method of wavefunction matching for calculating the
scattering wavefunction in the entire tunnel junction can be applied.
Experimentally, ballistic transport is observed for small and
defect-free samples.

The simplest approach to tunnel magneto-resistance traces back to the work
of Julli\`{e}re \cite{Julliere75}, where TMR is related to the polarization
of the junction defined via spin-dependent \textit{numbers of states} at the Fermi
energies $E_{\mathrm{F}}$ in the leads.
Maekawa and Gafvert \cite{Maekawa82} modified this model by defining the polarization
in terms of the \textit{densities of states} at $E_{\mathrm{F}}$.

In the Landauer-B\"uttiker theory \cite{Buettiker85}, the conductance
is proportional to the transmission probability of the scattering
channels (``conductance by transmission''), $G \propto
T(E_{\mathrm{F}})$. For a tunnel junction, the scattering channels are
the eigenstates of the leads in absence of the tunnel barrier, i.\,e.,
the Bloch states.  This approach can be applied in both sophisticated
electronic-structure calculations or in model calculations. For
example, Slonczewski assumed free electrons in the ferromagnetic leads
and a step-shaped tunnel barrier (as sketched in figure~\ref{fig:tunbarr})
\cite{Slonczewski89}.

The present work goes beyond Slonczewski's model, in that it takes
into account the nonzero potential in the leads. Applying the NFE
model, the occurrence of band gaps in the electronic structure has a
profound effect on the electronic transport. For the tunnel barrier,
we stick with the step shape, being aware that a differently shaped
barrier will lead to quantitatively different conductances
\cite{Mazin01}. The qualitative picture, however, will still be valid
\cite{Probst02}.  Although being computed in the framework of the
Landauer-B\"uttiker theory, the present results can also be interpreted
in terms of the densities of states.

\section{Model for the tunnel junction}
\label{sec:tunnel-junction}
The one-dimensional tunnel junction consists of three regions: the
left lead $\Lsys = \{ x | x < -d /2 \}$, the barrier $\Bsys = \{ x |
-d / 2 \leq x \leq d/ 2\}$, and the right lead $\Rsys = \{ x | d / 2 <
x \}$.  To keep the model as simple as possible, the leads were chosen
identical, i.\,e., identical lattice spacing and potential (This
restriction can, however, be relaxed in the \textsc{Mathematica}\texttrademark
\, notebooks \cite{Mathematica}). Within the leads, atoms are positioned
at $x = -(d / 2 + n a)$ and $x = +(d / 2 + n a)$ for \Lsys\ and \Rsys,
respectively (with $n$ semi-positive integer).  The lattice spacing
$a$ serves as the unit of length.

In accordance with the above spatial decomposition, the potential of
the tunnel junction consists also of three parts
(figure~\ref{fig:potential}), namely $V^{\Lsys}$, $V^{\Bsys}$, and
$V^{\Rsys}$. The barrier potential $V^{\Bsys}(x)$ is assumed
constant, with value $V_{\mathrm{barr}} > 0$. In principle, any other
shape could be used \cite{Ando87}. In the leads, the potential is
periodic, i.\,e., $V^{\Lsys}(x) = V^{\Lsys}(x - a)$ and $V^{\Rsys}(x)
= V^{\Rsys}(x + a)$.

\section{Calculation of the electronic structures}
\label{sec:electronic-structure}

\subsection{Electronic structure of the leads}
\label{sec:electr-struct-leads}
According to the Landauer-B\"uttiker theory, one needs to compute the
scattering channels, i.\,e., the electronic states of the leads in
absence of the tunnel barrier. Thus, the lead potential extends over
the whole $x$-axis. 

The Schr\"odinger equation $H \psi = E \psi$ in atomic units \cite{Units2} reads 
\begin{equation}
  \left[
    \frac{1}{2}
    \frac{\mathrm{d}^{2}}{\mathrm{d}x^{2}}
    +
    V(x)
  \right]
  \psi(x)
  =
  E \psi(x).
  \label{eq:2}
\end{equation}
Expanding the periodic potential into a Fourier series,
\begin{equation}
  V(x)
  =
  \sum_{q \in \mathbb{Q}}
  V_{q}
  \mathrm{e}^{\mathrm{i} q x},
  \label{eq:1}
\end{equation}
the requirement of periodicity [$V(x) = V(x + n a)$, $n$ integer]
restricts $q$ immediately to the reciprocal lattice $\mathbb{Q}$,
i.\,e., $q \in \mathbb{Q} = \left\{ n g | n\
\mathrm{integer}, g = 2\pi / a \right\}$. The mean value $V_{0}$ of
the potential serves as origin of the energy scale and, hence, is set
to zero ($V_{0} = 0$).  Assuming further a real $V(x)$ and the mirror
symmetry $V(x) = V(-x)$, one has $V_{-q} = V_{q}\ \forall q \in
\mathbb{Q}$.  For free electrons, $V(x)$ vanishes identically ($V_{q}
= 0\ \forall q \in \mathbb{Q}$).

For a periodic potential, the eigenstates $\psi$ of the Hamiltonian
$H$ fulfill Floquet's theorem \cite{Cottey71}.
In analogy to (equation~\ref{eq:1}), the Bloch states are as well expanded into a Fourier
series,
\begin{equation}
  \psi_{k}(x)
  =
  \mathrm{e}^{\mathrm{i} k x}
  \sum_{q \in \mathbb{Q}}
  c_{q}(k)
  \mathrm{e}^{\mathrm{i} q x}.
  \label{eq:3}
\end{equation}

In the following, the Schr\"odinger equation is solved by two methods,
which have in common that the differential equation is transformed
into an algebraic set.

\subsubsection{Computation of $E(k)$.}
\label{sec:computation-ek}
Inserting equation~\eqref{eq:1} and equation~\eqref{eq:3} into the Schr\"odinger equation
(\ref{eq:2}) gives the infinite algebraic set
\begin{equation}
  \left[
    \frac{1}{2}
    (q + k)^{2}
  \right]
  c_{q}(k)
  +
  \sum_{q' \in \mathbb{Q}}
  V_{q - q'}
  c_{q'}(k)
  =
  E(k)
  c_{q}(k),
  \quad
  \forall q \in \mathbb{Q},
  \label{GlSys}
\end{equation}
with $k \in ] -g/2, g/2]$.  Defining
\begin{equation}
  H_{q q'}(k)
  =
  \frac{1}{2} (q + k)^{2} \delta_{q q'}
  +
  V_{q - q'},
  \quad \forall q, q' \in \mathbb{Q},
  \label{eq:15}
\end{equation}
one obtains the compact form
\begin{equation}
  \sum_{q' \in \mathbb{Q}}
  H_{q q'}(k) c_{q'}(k)
  =
  E(k) c_{q}(k),
  \quad \forall q \in \mathbb{Q},
\end{equation}
which can be cast into matrix form, $H(k) \vec{c}(k) = E(k)
\vec{c}(k)$.  The Hamilton matrix $H(k)$ is square and hermitian,
resulting in real eigenvalues $E(k)$.

\subsubsection{Computation of $k(E)$.}
\label{sec:computation-ke}
In tunneling calculations, an electron is specified by its energy $E$.
Since the Bloch states in each lead are characterized by the
wavenumber $k^{\Lsys}$ and $k^{\Rsys}$, respectively, it has to be
assured that $E(k^{\Lsys}) = E(k^{\Rsys}) = E$. Alternatively, one
could solve for $k(E)$ in each lead, an obviously favorable method
\cite{Pendry69b}.

Rewriting equation~\eqref{GlSys} as
\begin{equation}
  \left(
    q - \sqrt{2 \left( E + V_0 \right)} + k
  \right)
  \left(
    q + \sqrt{2 \left( E + V_0 \right)} + k\right
  ) c_{q}
  +
  2
  \sum_{q' \in \mathbb{Q}\smallsetminus q} V_{q - q'} c_{q'}
  =
  0,
  \quad \forall q \in \mathbb{Q},  
\end{equation}
an algebraic set of double rank as in the $E(k)$ approach is
obtained by defining auxiliary expansion coefficients $w_{q}$:
\begin{equation}
\begin{split}
  (q - \sqrt{2 \left( E + V_0 \right)} + k) w_{q}
  +
  2 \sum_{q' \in \mathbb{Q}\smallsetminus q} V_{q - q'} c_{q'}
  & =
  0,
  \\
  -w_{q}
  +
  (q + \sqrt{2 \left( E + V_0 \right)} + k)
  c_{q}
  & =
  0.  
\end{split}
\label{eq:5}
\end{equation}
In matrix form, equation~\eqref{eq:5} reads
\begin{equation}
  \left(
    \begin{array}{cc}
      h_{11}(E) & h_{12}(E)
      \\
      h_{21}(E) & h_{22}(E)
    \end{array}
  \right)
  \left(
    \begin{array}{c}
      \vec{w}(E) \\
      \vec{c}(E)
    \end{array}
  \right)
  =
  k(E)
  \left(
    \begin{array}{c}
      \vec{w}(E) \\
      \vec{c}(E)
    \end{array}
  \right),
  \label{eq:16}
\end{equation}
where the Hamilton matrix $H(E)$ is a $2 \times 2$
supermatrix comprising submatrices $h_{ij}$ of rank $r_{\mathbb{Q}}$,
\begin{equation}
   \begin{split}
     h^{(11)}_{q q'} & = (q - \sqrt{2\left( E + V_0 \right)}) \,\delta_{q q'}, \qquad \forall q, q' \in \mathbb{Q},
     \\
     h^{(22)}_{q q'} & = (q + \sqrt{2\left( E + V_0 \right)}) \,\delta_{q q'},
     \\
     h^{(12)}_{q q'} & = 2\, V_{|q-q'|}\,(1-\delta_{q q'}),
     \\
     h^{(21)}_{q q'} & = - \delta_{q q'}.
   \end{split}
\end{equation}

Since $H(E)$ is not hermitian, its $2 r_{\mathbb{Q}}$ eigenvalues
$k^{(i)}(E)$ can be complex.  The dispersion relation $k(E)$ defines
then the complex band structure \cite{Heine63}. For the tunneling,
real $k^{(i)}$ are appropriate because only these belong to
square-integrable wavefunctions [$\psi^{(i)} \in L_{2}(-\infty,
+\infty)$ for $k^{(i)}$ real]. If, however, $\Im\, k^{(i)} > 0$
($\Im\, k^{(i)} < 0$), the eigenfunction $\psi^{(i)}$ decays
exponentially towards $+x$ ($-x$) \cite{Slater37}.

\section{Calculation of tunneling}
\label{sec:tunneling}
Having specified the electronic states in the three regions \Lsys,
\Bsys, and \Rsys, the wavefunction in the entire tunnel junction at a
given energy $E$ has to be determined.  This is achieved by requiring
continuity of the total wavefunction and of its first spatial
derivative at the two interfaces \Lsys-\Bsys\ at $x = -d/2$ and
\Bsys-\Rsys\ at $x = d/2$ (figure~\ref{fig:potential}).

While the Schr\"odinger equation determines the functional form of the
wavefunctions in the three regions, their actual shape has to be
determined by the boundary conditions. Since tunneling is viewed as
scattering of an incoming electron at the barrier
(figure~\ref{fig:tunbarr}), scattering boundary conditions have to be
applied. Note that
only electrons in occupied states of the source electrode can tunnel
into unoccupied states of the drain electrode, due to the Pauli
exclusion principle.

Assume for the following that the wavefunctions in the leads and in
the barrier at energy $E$ have been computed according to the
approaches introduced in section~\ref{sec:electronic-structure}. Since $E$ is
fixed, the explicit energy dependence will be dropped.

\subsection{Nonmagnetic leads}
\label{sec:non-magnetic-leads}

\subsubsection{Determination of the total wavefunction.}
In the present one-dimensional model, there is only a single Bloch
state in each lead with positive (or negative) velocity.  Thus, the
total wavefunction in \Lsys\ is superposed by the incoming Bloch state
$\psi_{\mathrm{i}}(x)$ (with velocity $v_{\mathrm{i}} > 0$) and the
reflected Bloch state $r \psi_{\mathrm{r}}(x)$ (with velocity
$v_{\mathrm{r}} = -v_{\mathrm{i}}$), $\psi_{\mathrm{i}}(x) + r
\psi_{\mathrm{r}}(x)$.  In \Rsys\ there is only the transmitted Bloch
state $t \psi_{\mathrm{t}}(x)$ with $v_{\mathrm{t}} > 0$
(figure~\ref{fig:tunbarr}).  Within the barrier \Bsys, two exponential
functions are taken, $A \exp(\kappa x) + B \exp(-\kappa x)$. The
matching conditions provide the probabilities for transmission through and
reflection at the barrier given by $|t|^{2}$ and $|r|^{2}$,
respectively.

Within Landauer-B\"uttiker theory, reservoirs contribute each incoming
scattering channel with the same current $j_{\mathrm{i}}$. Therefore,
either the Bloch states \cite{Slonczewski89} or the transmission
probabilities have to be normalized to unit current,
\begin{equation}
  T = |t|^2 \, \frac{j_{\mathrm{t}}}{j_{\mathrm{i}}},
  \label{eq:8}
\end{equation}
with the current $j_{\mathrm{t}}$ carried by $\psi_{\mathrm{t}}$
being calculated from $j = \Im(\psi^{\star} \, \frac{d}{dx} \psi)$.  The transmission
$T$ measures the transmitted current which is carried by
$\psi_{\mathrm{t}}$ upon feeding with $\psi_{\mathrm{i}}$.  Current
conservation implies $(1 - |r|^{2}) j_{\mathrm{i}} = |t|^2
j_{\mathrm{t}}$.

\subsubsection{Current and conductance.}
\label{sec:conductance}
Applying a bias voltage $V_{\mathrm{bias}}$ to the tunnel junction
shifts the Fermi energies of the electrodes, as is shown schematically
in figure~\ref{fig:bias}. For zero bias, the Fermi energies $E_{\mathrm{F}}$
of the electrodes \Lsys\ and \Rsys\ coincide ($E_{\mathrm{F}}^{\Lsys}
= E_{\mathrm{F}}^{\Rsys}$, figure~\ref{fig:bias}a).  For $V_{\mathrm{bias}}
> 0$, the potential $V^{\Rsys}(x)$ is shifted rigidly to lower
energies, which is achieved by taking as potential average $V_{0} =
-V_{\mathrm{bias}}$ instead of $V_{0} = 0$ (as being fixed for \Lsys).
Evidently, both the band structure and the Fermi energy of \Rsys\ are
shifted alike, $E^{\Rsys}(k) = E^{\Lsys}(k) - V_{\mathrm{bias}}$ and
$E_{\mathrm{F}}^{\Rsys} = E_{\mathrm{F}}^{\Lsys} - V_{\mathrm{bias}}$
(figure~\ref{fig:bias}b).  Hence, electrons can tunnel from \Lsys\ to \Rsys.
Since only electrons of occupied states in \Lsys\ (with energy $E <
E_{\mathrm{F}}^{\Lsys}$) can tunnel into unoccupied electronic states
in \Rsys ($E_{\mathrm{F}}^{\Lsys} - V_{\mathrm{bias}} < E$), an
energy window opens up in which tunneling can take place ($E \in
[E_{\mathrm{F}}^{\Lsys} - V_{\mathrm{bias}}, E_{\mathrm{F}}^{\Lsys}]$,
cf. the orange areas in figure~\ref{fig:bias}).

The total current $I$ through the tunnel junction, which flows from
the source electrode \Lsys\ to the drain electrode \Rsys, can be
expressed for small $V_{\mathrm{bias}}$ (linear response) as $I = G\,
V_{\mathrm{bias}}$.  Within Landauer-B\"uttiker theory, it is given by
the sum over all transmission probabilities $T(E)$ in the energy window
$[E_{\mathrm{F}}^{\Lsys} - V_{\mathrm{bias}},
E_{\mathrm{F}}^{\Lsys}]$,
\begin{equation}
  I
  =
  G_{0}
  \int_{E_{\mathrm{F}}^{\Lsys} - V_{\mathrm{bias}}}^{E_{\mathrm{F}}^{\Lsys}}
  T(E)
  \,\mathrm{d}E.
  \label{eq:10}
\end{equation}
In general, $T(E)$ implies a sum over the transmission probabilities
of all incident Bloch states $\psi_{\mathrm{i}}$ and all transmitted
Bloch states $\psi_{\mathrm{t}}$. In the present one-dimensional
model, with a single electronic state available in each electrode,
$T(E)$ reduces to a single term [cf equation \eqref{eq:8}].  The quantum of
conductance $G_{0}$ is $4 \pi$ in atomic units ($2 e^{2} / h$), where
the factor $2$ comes from the spin degeneracy.  In experiments, the differential conductance
$\mathrm{d}I / \mathrm{d}V_{\mathrm{bias}}$ is typically recorded,
which is computed numerically from equation~\eqref{eq:10}.

The conductance $G$ of the tunnel junction,
\begin{equation}
  G
  =
  \frac{G_{0}}{V_{\mathrm{bias}}}
  \int_{E_{\mathrm{F}}^{\Lsys} - V_{\mathrm{bias}}}^{E_{\mathrm{F}}^{\Lsys}}
  T(E)
  \,\mathrm{d}E,
  \label{eq:11}
\end{equation}
depends on details of the tunnel junction via the transmission $T(E)$,
e.\,g., on the electrode band structures and
properties of the tunnel barrier. 

\subsection{Ferromagnetic leads}
\label{sec:ferromagnetic-leads}

In nonmagnetic systems, each electronic state can be occupied by two
electrons, one with spin up, the other with spin down.  The
spin-dependent band structures of each lead coincide
[$E_{\pm}^{\Lsys}(k) = E(k)$ and $E_{\pm}^{\Rsys}(k) = E(k) -
V_{\mathrm{bias}}$] and the number of spin-up electrons equals the
number of spin-down electrons.  This degeneracy is lifted in magnetic
systems, and the spin-dependent band structures are rigidly shifted by
the exchange splitting $V_{\mathrm{ex}} > 0$ with respect to each
other, i.\,e., $E_{\pm}^{\Lsys}(k) = E(k) \mp V_{\mathrm{ex}} / 2$ and
$E_{\pm}^{\Rsys}(k) = E(k) - V_{\mathrm{bias}} \mp V_{\mathrm{ex}} /
2$ (\ref{fig:magtjbs}).  Since the electronic states are occupied up
to the Fermi energy $E_{\mathrm{F}}$, there are usually more spin-up 
electrons than spin-down electrons, which leads to the notation majority
(minority) electrons indicated by ``$+$'' (``$-$'').

Associating majority (minority) with spin-up (spin-down) electrons in
\Lsys\ \cite{Jonker04}, defines a net magnetic moment
$\vec{M}^{\Lsys}$ which is parallel to the magnetic moments of the
electrons, due to the surplus of majority electrons.  In a magnetic tunnel junction two
configurations remain for the right lead. Either the net magnetic moment 
$\vec{M}^{\Rsys}$ in \Rsys\, is parallel to that in \Lsys\,
($\vec{M}^{\Lsys} \parallel \vec{M}^{\Rsys}$) which is called P configuration or anti-parallel
($\vec{M}^{\Lsys} \parallel -\vec{M}^{\Rsys}$) which is called AP configuration, respectively.

In experiment, the configurations can be switched by
an external mag Because the spin is conserved during the tunneling process, the total
current in each configuration is given by the sum of the two
spin-resolved currents (``two current model''). In P configuration,
with electrons tunneling from majority (minority) states into majority
(minority) states, one has the upper situation of figure~\ref{fig:magtjconf}
\begin{equation}
  \label{eq:Ip}
  I_{\mathrm{P}}
  =
  I_{++}
  +
  I_{--},
\end{equation}
and accordingly for the AP configuration, with majority and minority
interchanged in \Rsys\ (bottom in figure~\ref{fig:magtjconf}),
\begin{equation}
  \label{eq:Iap}
  I_{\mathrm{AP}}
  =
  I_{+-}
  +
  I_{-+}.
\end{equation}
The four spin-dependent currents are related to the conductance
contributions $G_{\mathrm{P}} = G_{++} + G_{--}$ and $G_{\mathrm{AP}}
= G_{+-} + G_{-+}$.  The quantum of conductance $G_{0}$ is now $2 \pi$
(i.\,e., $e^{2} / h$). For a nonmagnetic junction, all spin-resolved currents are
identical ($I_{++} = I_{--} = I_{-}+ = I_{-+}$).

In typical tunnel junctions in P configuration, one of the
spin-resolved currents exceeds the other by far (for instance, $I_{++}
\gg I_{--}$ in figure~\ref{fig:magtjconf}), whereas in AP configuration,
both currents are of almost the same size ($I_{+-} \approx I_{-+}$).
The tunnel magneto-resistance (TMR) $\delta$ is a measure for the
spin-dependent transport and defined here by the asymmetry of
$I_{\mathrm{P}}$ and $I_{\mathrm{AP}}$,
\begin{equation}
  \delta
  =
  \frac{I_{\mathrm{P}} - I_{\mathrm{AP}}}{I_{\mathrm{P}} + I_{\mathrm{AP}}}
  =
  \frac{G_{\mathrm{P}} - G_{\mathrm{AP}}}{G_{\mathrm{P}} + G_{\mathrm{AP}}}.
  \label{eq:28}
\end{equation}

The tunneling calculations for ferromagnetic leads proceed as those
for nonmagnetic leads but with the band structures shifted by $\mp
V_{\mathrm{ex}} / 2$, giving rise to the four currents $I_{\pm\pm}$.
Eventually, $I_{\mathrm{P}}$, $I_{\mathrm{AP}}$, and $\delta$ are
computed from the latter.

\section{Results and discussion}
\label{sec:results}
The theory of tunneling, as described in the preceding sections, was
implemented in a set of \textsc{Mathematica}\texttrademark \, notebooks
\cite{Mathematica}.  Because the results for a magnetic tunnel
junction (subsection~\ref{sec:ferromagnetic-leads-1}) can easily be understood
from those for a nonmagnetic one, the latter are presented and
discussed first. For the following, the basis set is $\mathbb{Q} = \{
-3g, -2g, \ldots, +2g \}$ ($r_{\mathbb{Q}} = 6$). The Fourier
coefficients $V_{q}$ of the lead potentials are chosen as $V_{0} =
0$\,H, $V_{g} = 0.012$\,H, $V_{2g} = 0.008$\,H, and $V_{3g} =
0.004$\,H.

\subsection{Nonmagnetic leads}
\label{sec:non-magnetic-leads-1}

\subsubsection{Nearly-free \textit{vs} free electrons.}
\label{sec:nearly-free-textitvs}
The Hamilton matrix for free electrons is diagonal ($V_{q} = 0$), and
the Bloch states become pure plane waves, $\phi^{(q)} =
\exp[\mathrm{i} (k + q) x]$, with band structure $E^{(q)}(k) = (k +
q)^{2} / 2$ and velocities $v^{(q)}(k) = k + q$. The density of states decreases with
energy, $D(E) = 2 \pi\, \mathrm{d}k(E)/\mathrm{d}E = \pi \sqrt{2 / E}$.

The nonzero potential in the NFE case results in band gaps at $k = 0$
and $k = \pm g/2$. The band gaps at $k = \pm g/2$ are between band~1
(lowest in energy, black in figure~\ref{fig:elstru}a and \ref{fig:elstru}e) and band~2
(orange) and show up between $0.112$\,H and $0.137$\,H. The first gap
at $k = 0$ appears at $0.492$\,H between bands~2 and 3. The band
widths are much larger than the band gaps. For band~1, for instance,
the band width is $E^{(1)}(g/2) - E^{(1)}(0) = 0.112\,\mathrm{H} +
0.006\,\mathrm{H} = 0.118\,\mathrm{H}$, whereas the band gap is
$E^{(2)}(g/2) - E^{(1)}(g/2) = 0.025$\,H.  Thus, the electrons behave
as \emph{nearly free} (NF), that is, the electronic properties differ
from those of free electrons in the small energy ranges at the band
gaps.

The velocities $v^{(i)}(k)=\mathrm{d}E^{(i)}(k)/\mathrm{d}k$ for NF electrons (figure~\ref{fig:elstru}b)
which are linear in a large part of the BZ, vanish at the band edges
($k = \pm g/2$ and $k = 0$). Further, they change sign at $k = 0$.
Since electrons with zero velocity cannot contribute to the current, a
considerable effect on the tunnel current is expected at the band
gaps. The DOS $D(E)$ (figure~\ref{fig:elstru}e) also resembles that of free
electrons but becomes singular at the band edges. These Van-Hove
singularities will also show up in the tunnel current. Considering the
NFE Bloch states, the expansion coefficient $c_{0}^{(1)}$ of band~1
dominates in a large part of the BZ, that is, the wavefunction is to a
very good approximation equal to the single plane wave with $q = 0$
(figure~\ref{fig:elstru}c). At $k = \pm g/2$, other plane waves become mixed
in: $|c_{-g}^{(1)}| > 0$ at $k = +g/2$ and $|c_{g}^{(1)}| > 0$ at $k =
-g/2$. Band~2 shows the same qualitative behaviour, but in addition
mixing at $k = 0$. In summary, differences between free and
nearly-free electrons in the leads show up close to band gaps.

In order to work out how the tunneling is modified with respect to
that of free electrons, the transmissions $T(E)$ \textit{vs} barrier
width $d$ are compared for energies close to and well below the band
gap at $k = g/2$.  For $d = 0$, the transmission is perfect in all
cases [$T(E) = 1$] because the incoming Bloch states are not
reflected at the barrier (figure~\ref{fig:transmission}a).  With
increasing $d$, $T(E)$ decreases due to the decaying wavefunctions
in \Bsys. For $E = 0.112$\,H, i.\,e., close to the band edge, the
transmission for free electrons is significantly larger than for NF
electrons, as can be explained by their larger velocity.  On the
contrary, there is no apparent difference for $E$ well below the
band edge (here: $E = 0.050$\,H) because in both cases the Bloch
states comprise a single plane wave. In summary, band gaps have a
pronounced effect on the transmission and show also up in the bias
dependence of the current.

\subsubsection{Bias dependence of tunneling.}
\label{sec:bias-depend-tunn} Before turning to current and
conductance, the dependence of the transmission on bias voltage and
barrier width is addressed for fixed energy
(figure~\ref{fig:transmission}b). The overall smooth shape is disturbed by
sharp minima at about $0.013$\,H and $0.4$\,H bias
which are related to the band gaps in \Rsys.  The monotonous decay
with barrier width $d$ is consistent with that in
figure~\ref{fig:transmission}.

At first glance, one would expect that a current $I$ through a tunnel
junction increases with bias voltage, in accordance with Ohm's law.
This behaviour can indeed be observed in figure~\ref{fig:biasdep}a, but only
as a general trend. For bias below $0.1$\,H and at about $0.4$\,H
ample deviations occur, with $I$ showing even a decrease.  These
features show up more pronounced in the differential conductance
$\mathrm{d}I / \mathrm{d}V$ (figure~\ref{fig:biasdep}b), where the linear
dependence of $I$ on bias can be clearly resolved between $0.10$\,H
and $0.35$\,H. The conductance drops rapidly with bias but remains
almost constant for $V_{\mathrm{bias}} > 0.1$\,H
(figure~\ref{fig:biasdep}c).

The aforementioned features can be discussed in terms of band
structures and transmission probabilities. However, a simpler access
is provided by the number of states within the energy window of
tunneling $[ E_{\mathrm{F}}^{\Lsys} - V_{\mathrm{bias}},
E_{\mathrm{F}}^{\Lsys} ]$, an approach to be viewed as an extension of
the Maekawa-Gafvert model to nonzero bias
(section~\ref{sec:basic-theor-ball}). The number
$N_{\mathrm{i}}^{\Lsys}(V_{\mathrm{bias}})$ of incident states in
\Lsys\ is
\begin{equation}
  N_{\mathrm{i}}^{\Lsys}(V_{\mathrm{bias}})
  =
  \frac{1}{2}
  \int_{E_{\mathrm{F}}^{\Lsys} - V_{\mathrm{bias}}}^{E_{\mathrm{F}}^{\Lsys}}
  D^{\Lsys}(E)
  \,\mathrm{d}E,
  \label{eq:18}
\end{equation}
where the factor $1/2$ takes into account that only half of the Bloch
states show positive velocity. For $E_{\mathrm{F}}^{\Lsys}$ within
band~1 (at $0.1$\,H), $N_{\mathrm{i}}^{\Lsys}$ increases monotonously
until $V_{\mathrm{bias}} = E_{\mathrm{F}}^{\Lsys} -
E_{\mathrm{bot}}^{\Lsys}$, i.\,e., when the bottom of band~1 is
reached [$E_{\mathrm{bot}}^{\Lsys} = E^{(1)}(0)$;
figure~\ref{fig:biasdep}d]. Because $N_{\mathrm{i}}^{\Lsys}$ remains
constant for larger bias (figure~\ref{fig:bias} and figure~\ref{fig:elstru}), it
is sufficient to restrict the integration in equation~\eqref{eq:18} to
$[E_{\mathrm{min}}^{\Lsys}, E_{\mathrm{F}}^{\Lsys}]$, with
$E_{\mathrm{min}}^{\Lsys} = \max(E_{\mathrm{bot}}^{\Lsys},
E_{\mathrm{F}}^{\Lsys} - V_{\mathrm{bias}})$.

The number $N_{\mathrm{t}}^{\Rsys}(V_{\mathrm{bias}})$ of transmitted
states in \Rsys\ depends on the energy range in which incident
electrons are provided by \Lsys. Thus,
\begin{equation}
  N_{\mathrm{t}}^{\Rsys}(V_{\mathrm{bias}})
  =
  \frac{1}{2}
  \int_{E_{\mathrm{min}}^{\Lsys}}^{E_{\mathrm{F}}^{\Lsys}}
  D^{\Rsys}(E)
  \,\mathrm{d}E.
  \label{eq:19}
\end{equation}
As is evident from figure~\ref{fig:biasdep}d, the general shape of
$N_{\mathrm{t}}^{\Rsys}$ consists of an increase for bias up to about
$0.13$\,H and a slight decrease for larger bias. But on top of this,
it shows a richer structure than $N_{\mathrm{i}}^{\Lsys}$. Upon
increasing $V_{\mathrm{bias}}$, a kink appears either when a band edge
enters the energy windows (at $E_{\mathrm{F}}^{\Lsys}$) or when it
leaves the window (at $E_{\mathrm{min}}^{\Lsys}$). This is supported
by considering the DOS of \Rsys\ at the boundaries of the energy
window from (figure~\ref{fig:biasdep}e). At the upper boundary
[$D^{\Rsys}(E_{\mathrm{F}}^{\Lsys})$, black in figure~\ref{fig:biasdep}e], a
band gap produces a decrease of the current since there are less
transmitted states available.  The lower boundary
[$D^{\Rsys}(E_{\mathrm{min}}^{\Lsys})$, orange], however, shows much
less effect. The small kink in $I$ and $G$ at $0.1$\,H is due to
$N_{\mathrm{i}}^{\Lsys}$, because it appears just when
$E_{\mathrm{F}}^{\Lsys} - E_{\mathrm{bot}}^{\Lsys} =
V_{\mathrm{bias}}$.

In summary, the structures in the bias dependence of both the current
and the conductance can be explained by the density of states at the
boundaries of the tunnel-energy window and by the number of incoming
and transmitted states. The actual current, however, is further
influenced by the tunneling probability which appears to be of minor
importance due to its smooth shape.


\subsection{Ferromagnetic leads}
\label{sec:ferromagnetic-leads-1}

Having investigated the transport properties of tunnel junctions with
nonmagnetic leads, we now turn to junctions with ferromagnetic leads.
With regard to subsection~\ref{sec:ferromagnetic-leads}, the spin-dependent
currents $I_{\pm\pm}$ and conductances $G_{\pm\pm}$ were obtained from
calculations for nonmagnetic leads in which the band structures
$E_{\pm}(k)$ were rigidly shifted by the exchange splitting $\mp
V_{\mathrm{ex}} / 2$ (chosen as $0.03$\,H in the following).

Considering the conductances for the P configuration (figure~\ref{fig:tmr}a),
the arbitrarily chosen exchange splitting $V_{\mathrm{ex}}$ causes a bias dependence of the
conductance contribution $G_{--}$ which is very similar to the nonmagnetic
case (figure~\ref{fig:biasdep}b). The conductance contribution $G_{++}$,
however, is zero for small bias until the bias bridges the gap to the
top of band 1 in \Lsys. For larger bias, $G_{++}$ is finite
but always much smaller than $G_{--}$. Consequently, $G_\mathrm{P}$ is mainly
determined by $G_{--}$ (figure~\ref{fig:tmr}a).

In comparison to the P configuration, the conductances in AP configuration appear slightly
modified. $G_{-+}$ is zero until $V_{\mathrm{bias}}$ bridges the gap to the bottom of band
2 in \Rsys. It increases until a maximum is reached, drops rapidly,
and remains almost constant for $V_{\mathrm{bias}} > 0.1$\,H until the next gap occurs.
$G_{-+}$(AP) is similar to $G_{--}$(P) but the corresponding extrema are shifted to lower energies by
$V_{\mathrm{ex}}$. $G_{+-}$(AP) is similar to $G_{++}$(P) with the same onset at bias voltages
reaching the top of band 1 in \Lsys. $G_{\mathrm{AP}}$ is determined by $G_{-+}$.

The tunnel magneto-resistance $\delta$ (equation~\ref{eq:28}) shows a pronounced bias dependence
which is related to the bandstructure, especially to the position of the bandgaps of the leads
(figure~\ref{fig:tmr}c). First of all, a zero-bias anomaly shows up which
is caused by the half-metallic behaviour of the leads. This is ---in contrast to other explanations
\cite{Zhang97b}--- a pure electronic effect.
In general, $\delta$ decays with increasing $V_{\mathrm{bias}}$ except the influence
at the bandgaps. As discussed above (figure~\ref{fig:biasdep}),
a bandgap in $\Rsys$ at the upper boundary of the energy window causes a decrease
of the conductance. Minima in $G_\mathrm{P}$ manifest themselves as minima in $\delta$.
Conductance minima at $G_{\mathrm{AP}}\, (G_{-+})$ cause maxima at $V_{\mathrm{bias}}= 0.036$\,H and
$0.39$\,H. These maxima and the subsequent minima are separated by
$V_{\mathrm{ex}}$.

Concerning the thickness dependence of the tunneling magneto-resistance: the absolute
values of $\delta$ increase. Hence, the characteristic features of
$\delta$ become more pronounced. The reason for this behavior is
related to the fact that $G_\mathrm{P} + G_{\mathrm{AP}}$ drops faster with increasing barrier thickness than
$G_\mathrm{P} - G_{\mathrm{AP}}$.

\section{Concluding remarks}
\label{sec:concluding-remarks}

The main purpose of this paper was to demonstrate the role of a
realistic bandstructure of the leads, described within a NFE model,
for the current-voltage characteristics of tunnel junctions. The
consideration was extended to ferromagnetic junctions in order to
discuss the bias dependence of the tunnel magneto-resistance. In
difference to a free-electron model a pronounced TMR effect was
obtained. The TMR reveals a zero-bias anomaly of electronic origin
caused by the half-metallic behaviour of the leads. Driven by the
position of the band gaps in the drain electrode, with respect to
the Fermi energy in the source electrode, pronounced positive and
negative TMR values occur.

As further extensions to the presented theory, one could conceive to discuss
the spin polarizations defined in terms of the number of electrons in
the leads and of the spin-resolved currents $I_{\pm\pm}$. Contact to
the popular Julli\`{e}re and the Maekawa-Gafvert models
\cite{Julliere75,Maekawa82} can be made by trying to explain the TMR
$\delta$ by these spin polarizations. However, one has to be aware
that a possible agreement does not hold in general for
three-dimensional models because in one dimension the conductance is
related simply to the number of states in the leads. Another issue could involve
different leads which result in a nonsymmetric current-voltage
characteristics [$I(V_{\mathrm{bias}}) \not= -I(-V_{\mathrm{bias}})$].

\newpage
%
%
\bibliographystyle{plain}

\begin{thebibliography}{widest entry}
\bibitem{Units2}
Atomic Hartree units are used, $e = \hbar = m = 1$, $c \approx 137.036$.
  $1~\mathrm{H} \approx 27.12~\mathrm{eV}$.
\end{thebibliography}

%
%
\clearpage
\section*{Figure captions}

\begin{figure}[ht]
  \centering
  \caption{Sketch of a tunnel junction with ferromagnetic leads
    (\Lsys, \Rsys) and a step-shaped barrier (\Bsys). An electron
    incident from the left (with current~$j_{\mathrm{i}}$) is
    reflected (current $j_{\mathrm{r}}$) and transmitted through
    \Bsys\ into \Rsys\ ($j_{\mathrm{t}}$). In the leads, the potential
    depends on the spin of the electron: for a spin-up electron it is
    lower ($\uparrow$, dashed red line) than for a spin-down electron
    ($\downarrow$, solid red line).}
  \label{fig:tunbarr}
\end{figure}

\begin{figure}[ht]
  \centering
  \caption{Potential $V(x)$ of the tunnel junction.
    For the barrier \Bsys, a step-shaped potential with height
    $V_{\mathrm{barr}}$ and width $d$ is assumed. The potential in the
    leads \Lsys\ and \Rsys\ is periodic with the lattice spacing $a$.
    The interfaces between the three regions are at $x = \pm d/2$.}
  \label{fig:potential}
\end{figure}

\begin{figure}[ht]
  \centering
  \caption{Sketches of the electronic structure of a tunnel junction
    with a bias voltage applied, as depicted schematically by the
    density of states (black areas) for the leads \Lsys\ and \Rsys.
    For zero bias, the Fermi energies $E_{\mathrm{F}}$ coincide (a),
    whereas for positive bias $E_{\mathrm{F}}^{\Rsys}$ is lowered to
    $E_{\mathrm{F}}^{\Lsys} - V_{\mathrm{bias}}$ in \Rsys, opening up
    the energy window of tunneling [orange area in (b--d)]. The size
    of the energy window increases with $V_{\mathrm{bias}}$ (b) until
    its lower boundary $E_{\mathrm{F}}^{\Lsys} - V_{\mathrm{bias}}$
    reaches the bottom $E_{\mathrm{bot}}^{\Lsys}$ of the \Lsys-bands
    (c, d).  The barrier is denoted \Bsys.}
  \label{fig:bias}
\end{figure}

\begin{figure}[ht]
  \centering
  \caption{Spin-resolved band structures in the ferromagnetic leads
    of a tunnel junction in P configuration.  The majority-electron
    bands ($+$, black) are shifted by $V_{\mathrm{ex}} > 0$ with
    respect to the minority bands ($-$, orange) in both leads.  The
    junction is biased by $V_{\mathrm{bias}} > 0$.}
    \label{fig:magtjbs}
  \caption{Configurations of a tunnel junction with ferromagnetic leads.
    The lead magnetizations are either parallel ($\vec{M}^{\Lsys}
    \parallel \vec{M}^{\Rsys}$; P configuration, top) or anti-parallel
    ($\vec{M}^{\Lsys} \parallel -\vec{M}^{\Rsys}$; AP configuration,
    bottom). The spin-resolved currents $I_{\pm\pm}$ are visualized by
    horizontal arrows.}
  \label{fig:magtjconf}
\end{figure}

\begin{figure}[ht]
  \centering
  \caption{Electronic structure of a nonmagnetic lead,
    with a focus either on the $E(k)$ (left: a--d) or on the $k(E)$
    description (right: e--h).  (a) Bandstructure $E(k)$ of the lowest
    (black, band~1) and the second (orange, band~2) band in the first
    Brillouin zone, $k \in ]-g/2, g/2]$. (b) Velocity $v(k)$ of the
    corresponding Bloch states. (c, d) Fourier coefficients
    $c_{q}^{(i)}(k)$ of the Bloch states associated with band~1 (c)
    and 2 (d).  (e) Bandstructure, but in $k(E)$ representation.  (f)
    Density of states $D(E)$ for the bands shown in (a) and (d).  (g,
    h) Fourier coefficients of (c) and (d), but for $k \geq 0$ (g) and
    $k \leq 0$ (h).}
\label{fig:elstru}
\end{figure}

\begin{figure}[ht]
  \centering
  \caption{(a) Transmission $T$ through an unbiased tunnel junction with
    nonmagnetic leads ($V_{\mathrm{barr}} = 0.12$\,H). For energies
    close to ($E = 0.112$\,H; black lines) or well below ($E = 0.050$\,H;
    orange lines) the band gap at $k = g/2$ ($E = 0.112$\,H), $T(d)$ is
    plotted \textit{vs} barrier width $d$ for nearly-free (solid) and
    free (dotted) lead-electrons. (b) Transmission probability $T(E)$ \textit{vs} bias and barrier
    width $d$ at $E = 0.1$\,H.}
  \label{fig:transmission}
\end{figure}

\begin{figure}[ht]
  \centering
  \caption{Current-voltage characteristic of a tunnel junction with
    barrier width $d = 2.5$\,a and nonmagnetic leads.  (a-c) Total
    current $I$, differential conductance $\mathrm{d}I / \mathrm{d}
    V_{\mathrm{bias}}$, and conductance~$G$ \textit{vs} bias voltage.
    (d) Number of states incoming from \Lsys\ (black) and outgoing in
    \Rsys\ (orange). (e) Density of states of the right lead \Rsys\ at
    $E_{\mathrm{F}}^{\Lsys}$ (black) and at $E_{\mathrm{bot}}^{\Lsys}$
    (in units of the lattice constant).  The Fermi energy is $0.1$\,H.}
  \label{fig:biasdep}
\end{figure}

\begin{figure}[ht]
  \centering
  \caption{Spin-dependent tunneling for $E_{\mathrm{F}} = 0.1$\,H and
    exchange splitting $V_{\mathrm{ex}} = 0.03$\,H. (a) Spin-dependent
    conductances $G_{++}$ and $G_{--}$ for the P configuration
    \textit{vs} bias for barrier width $d = 2.5\,a$. Their sum is
    denoted $G_{\mathrm{P}}$. (b) As for (a)
    but for the AP configuration, with $G_{\mathrm{AP}} = G_{+-} +
    G_{-+}$. (c) Tunnel magneto-resistance $\delta$ obtained from the
    conductances of (a) and (b). (d) Dependence of the TMR $\delta$ on
    barrier width, as indicated by different line styles.}
  \label{fig:tmr}
\end{figure}

%
%
\pagestyle{empty}
\clearpage
\centering
\includegraphics[width = 0.98\columnwidth]{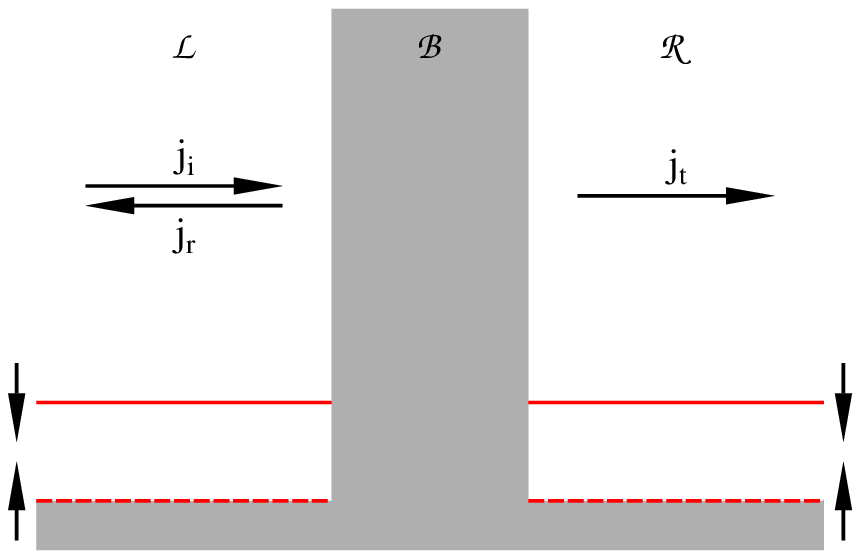}
\vfill Figure~\ref{fig:tunbarr}

\pagestyle{empty}
\clearpage
\centering
\includegraphics[angle=270,width = 0.98\columnwidth]{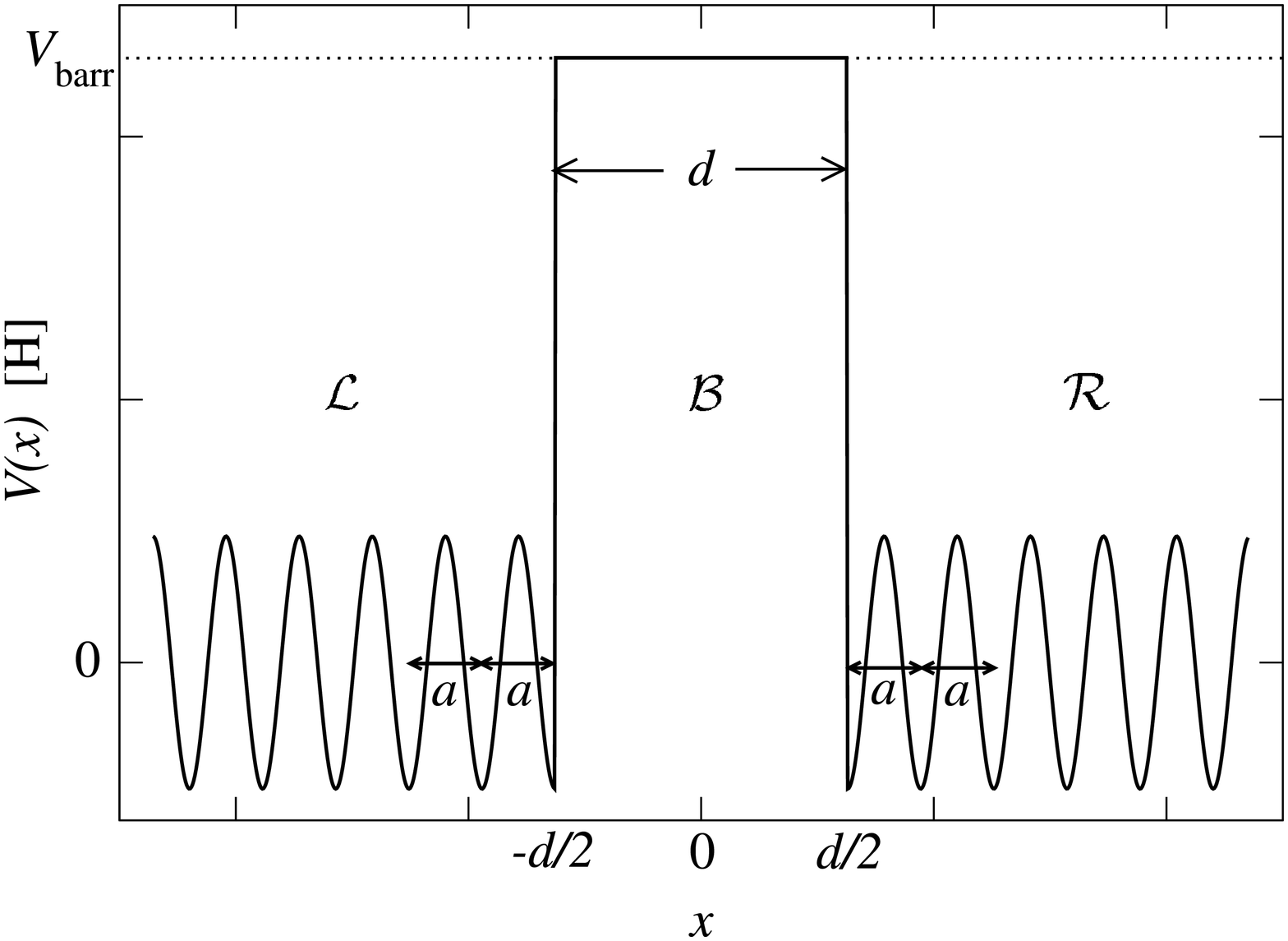}
\vfill Figure~\ref{fig:potential}

\pagestyle{empty}
\clearpage
\centering
\includegraphics[angle=270,width = 0.98\columnwidth]{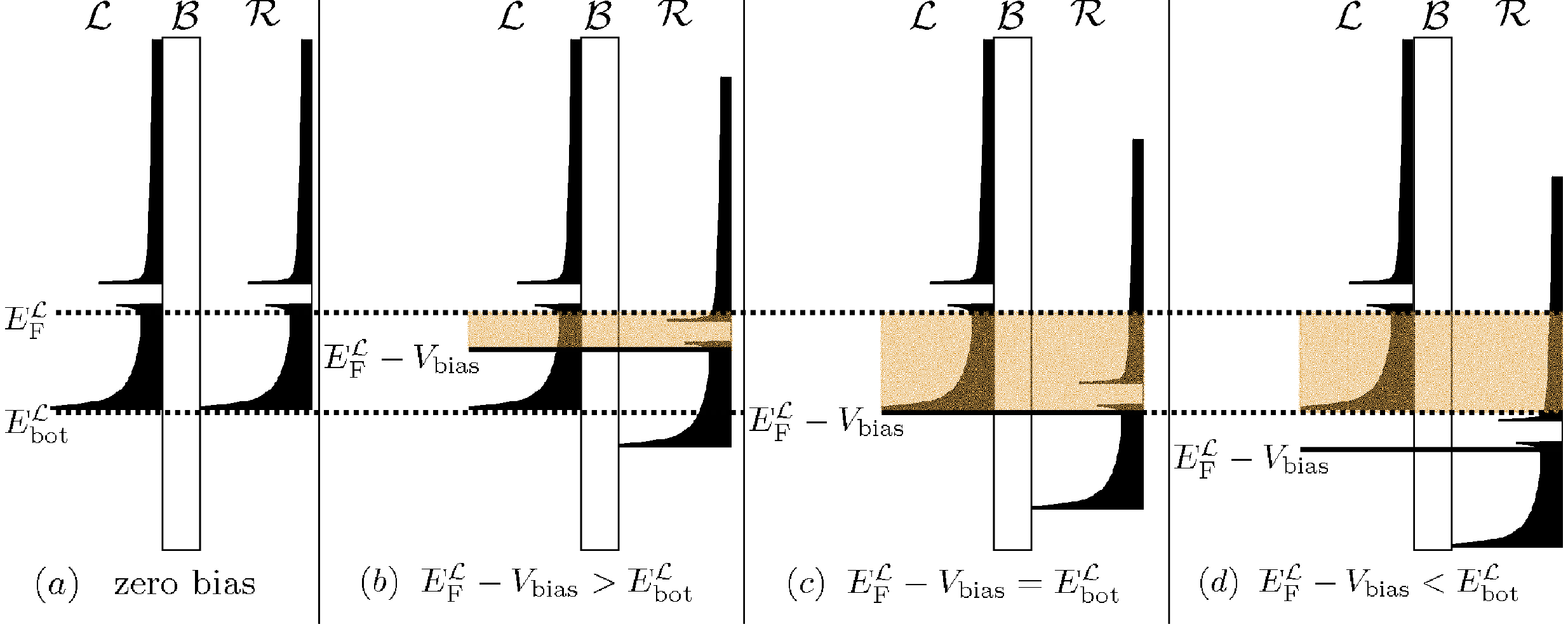}
\vfill Figure~\ref{fig:bias}

\pagestyle{empty}
\clearpage
\centering
\includegraphics[angle=270,width = 0.98\columnwidth]{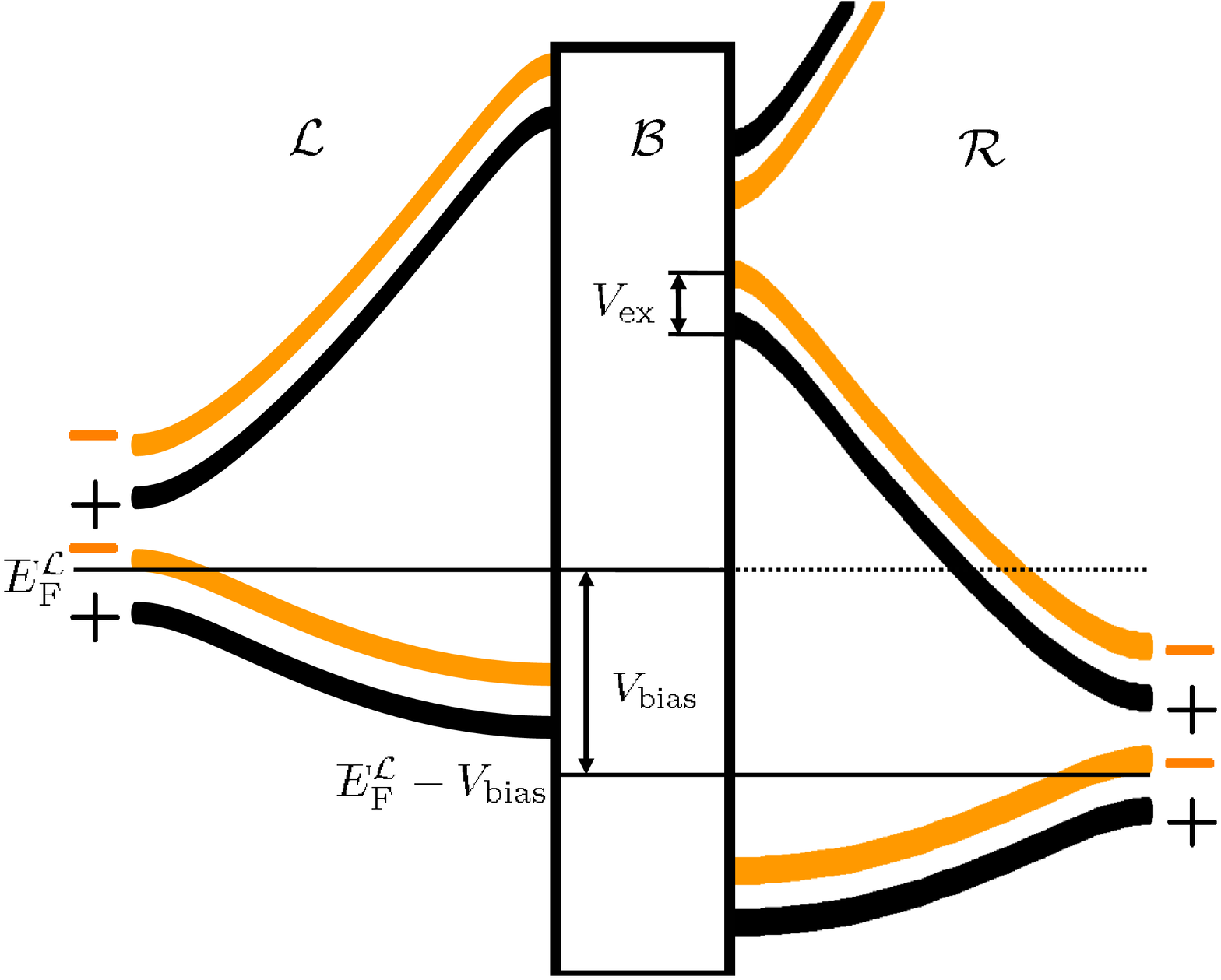}
\vfill Figure~\ref{fig:magtjbs}

\pagestyle{empty}
\clearpage
\centering
\includegraphics[angle=270,width = 0.98\columnwidth]{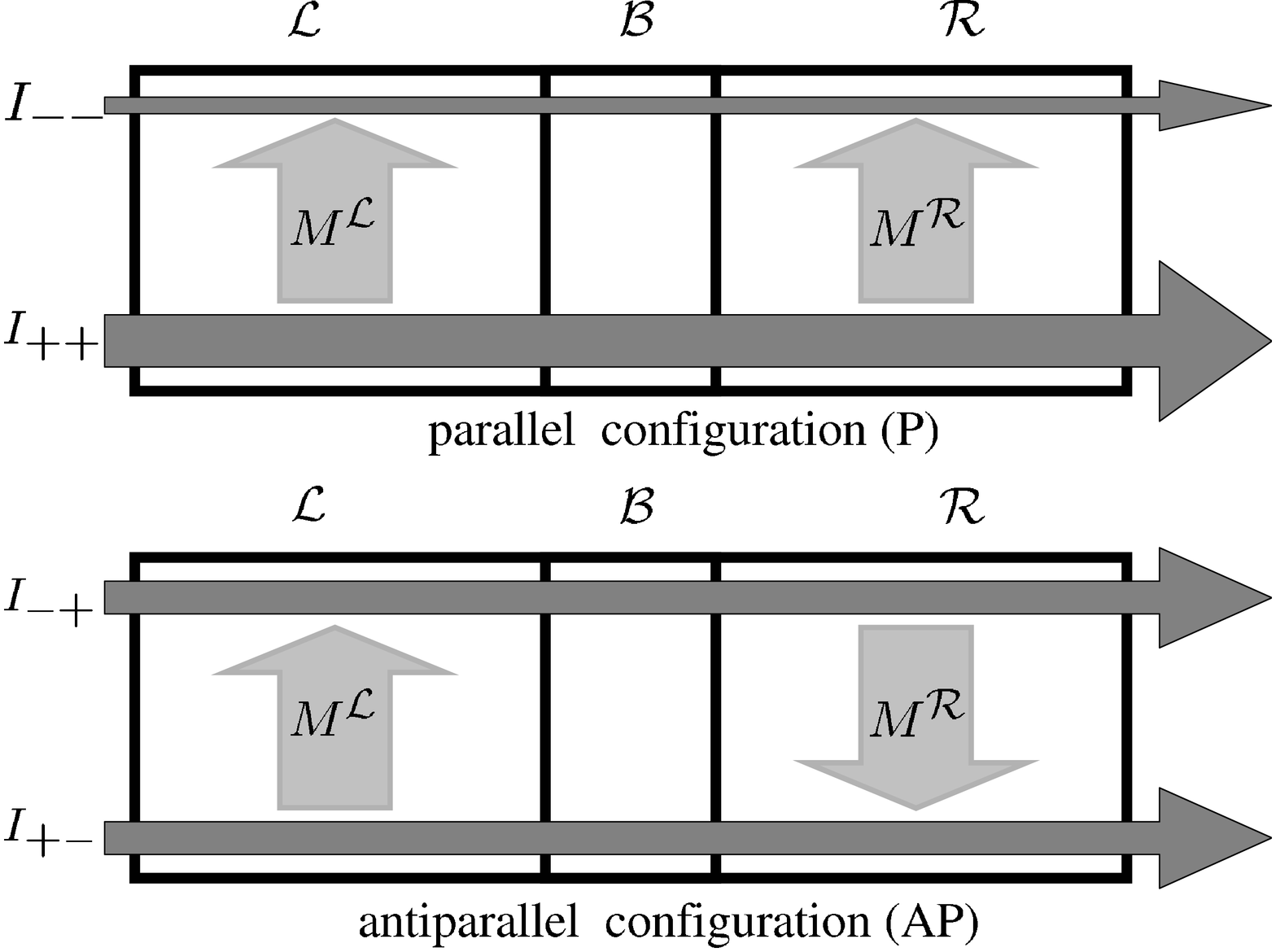}
\vfill Figure~\ref{fig:magtjconf}

\pagestyle{empty}
\clearpage
\centering
\includegraphics[width = 0.98\columnwidth]{figures/figure6}
\vfill Figure~\ref{fig:elstru}

\pagestyle{empty}
\clearpage
\centering
\includegraphics[width = 0.98\linewidth]{figures/figure7a}
\vfill Figure~\ref{fig:transmission}a

\pagestyle{empty}
\clearpage
\begin{figure}[b]
  \centering
  \includegraphics[angle =-90,width = 0.9\linewidth]{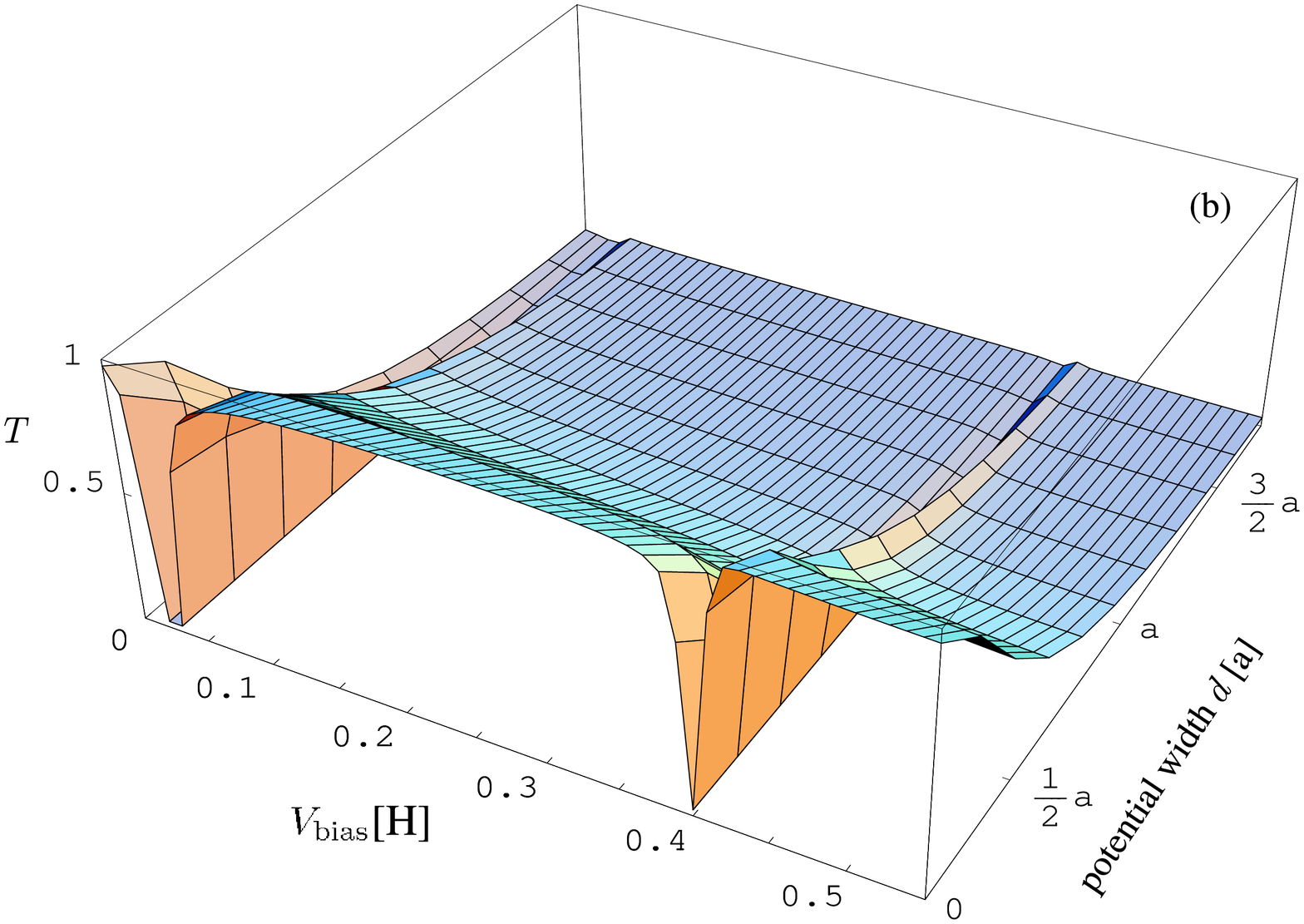}  
\end{figure}
\vfill Figure~\ref{fig:transmission}b


\pagestyle{empty}
\clearpage
\centering
\includegraphics[angle =0,width = 0.90\linewidth]{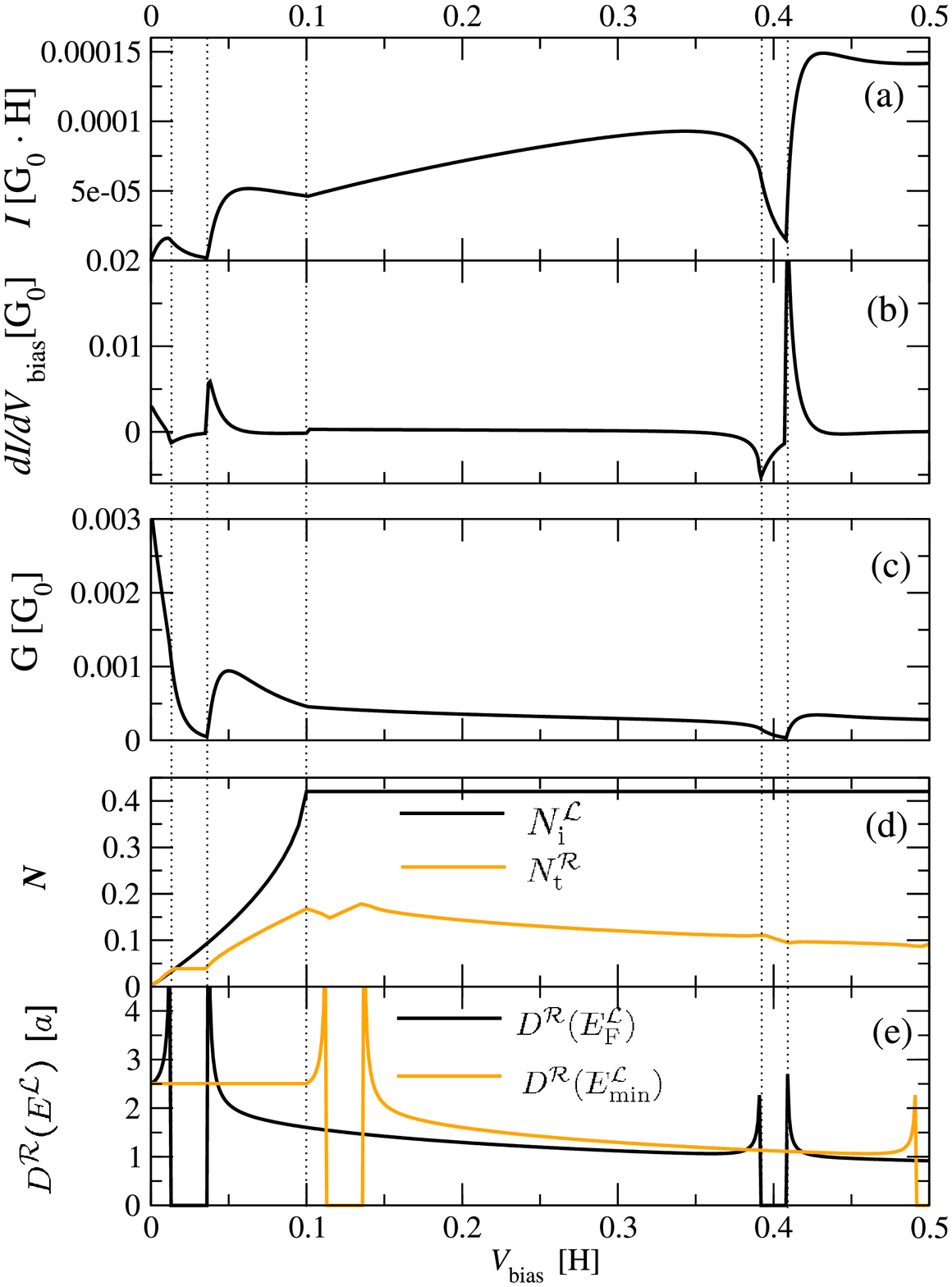}
\vfill Figure~\ref{fig:biasdep}

\pagestyle{empty}
\clearpage
\centering
\includegraphics[angle =0,width = 0.98\linewidth]{figures/figure9}
\vfill Figure~\ref{fig:tmr}

\end{document}